\documentstyle[12pt]{article}
\begin{document}

\author{P.\ M. Nadolsky, S.\ M.\ Troshin,
 N.\ E.\ Tyurin\\
Institute for High Energy Physics\\
142284 Protvino, Moscow Region, Russia}
\title{Near Future Perspectives of QCD Spin Studies\rm}
\date{}
\maketitle
\begin{abstract}
We consider the physics motivations and perspectives for
the study of spin phenomena at the future high  energy
accelerators. The possibilities to use
the already operating machines are also discussed.
It is emphasized that the present status of QCD spin studies
 necessarily requires wide range of spin measurements.
\end{abstract}
\tableofcontents
\section*{Introduction}
\addcontentsline{toc}{section}{Introduction}
The physics of spin effects in particle interactions at
large and small distances
provides valuable information on the fundamental properties of particles:
their wave functions,
short distance behavior of the lepton, quark and gluon
interactions, mechanisms of chiral symmetry breaking and
confinement.

Experiments play the leading role in study of spin
phenomena at present and provide a fuel for theoretical
analysis.
The experiments to study spin phenomena are foreseen at
almost all new accelerator facilities.
 By this time few new accelerators --- colliders,
have  been designed.  These  are  the
SSC, \index{SSC} proton--proton superconducting supercollider
with the beam energy 20 TeV under construction now and the
proton--proton large collider \index{LHC}
(LHC) \index{LHC} with the beam energy 8 TeV designed  at  CERN.
 The UNK \index{UNK}, a fixed target machine
with the energy 0.6 TeV at first stage is under construction at
IHEP, Serpukhov. Here a wide range of spin studies in hadronic reactions
is planned in experiments with internal polarized jet
target.

It is very essential for spin experiments that Siberian snake
concept makes the real possibility to have accelerated polarized
beams \cite{sib}.
There are plans to have accelerated polarized beam at
Fermilab Main Injector and Tevatron--Collider \cite{spin} and these studies
were commissioned by Fermilab. Collisions of polarized proton beam
with unpolarized
antiproton beam at $\sqrt{s}=2$ TeV  could be realized at
Tevatron--Collider and this will allow to test spin properties of
QCD at highest energy in various hadronic processes.

Acceleration
of polarized protons up to 2 TeV could be realized at SSC.

There are plans to have the polarized proton beams at
Relativistic Heavy Ions Collider (RHIC, BNL) \cite{bunce}.
Proposal on spin physics at RHIC \cite{rhic} has been
initially approved.

At DESY the $ep$--collider
(HERA)
\index{HERA} with the energy of electrons  35 GeV and
the proton energy 820 GeV  (most probably the  proton  beam  energy
will be raised up to 1 TeV) is already in operation.
Experiment HERMES devoted to
study spin structure functions in deep inelastic scattering
\cite{hermes} also has been approved. It will provide data
on the structure of nucleons and the tests of QCD, in particular, Bjorken
sum rule.

 At CERN the large electron--positron collider LEP with the energy
50$\times$50 GeV is operating. The LEP--200 project is underway.
Following the known EMC experiment the new SMC experiment repeated
the measurements
of the proton spin structure function performed earlier
by the EMC group at SPS but with better
accuracy.

The first linear electron--positron collider (SLC) started to
operate recently at SLAC with polarized beams. The first exciting
results on spin structure of neutron have been obtained.
Construction of the linear  electron--positron
colliders with the center--of--mass energy from 0.5  to  2
TeV is now being worked out also.

The  main  goal    of  these
facilities is exploration of the energy range,
characterized by the scale of 1 TeV.
It is  expected that
 the experiments would reveal a
spectrum of new phenomena  related to  the
 Higgs boson, new
gauge   bosons   and   supersymmetric     particles,
manifestation of the compositeness  of leptons and quarks and
possible mechanisms of mass generation.

Of course, there is a chance that some of new phenomena
will be found without measurements of the spin observables.
 However, the polarization measurements provide
additional opportunities to detect new physics.
 Moreover, these
measurements are absolutely necessary to study chiral
structure of the new particles couplings.

We review  here the possibilities to study
hadron dynamics with the help of spin effects as the closest
perspectives in the field.
 Also the perspectives related to electroweak interactions and new
physics are mentioned.

 \section{Spin Studies in Hadronic Reactions}
\subsection{Chiral Invariance and Spin Properties of QCD}
Nowadays Quantum Chromodynamics (QCD) is generally accepted as
a theory of strong interaction. The perturbative expansion based on the
asymptotic freedom of QCD allows one to calculate the observables in
hard processes and apply this theory to the world of particle
interactions.
The QCD lagrangian has the form
\begin{equation}
{\cal L}_{QCD}=
\bar{\psi }(x)(i\hat D-m)\psi (x)
-\frac{1}{4}\mbox{tr}(G_{\mu \nu }G^{\mu \nu }),
\label{4.22}
\end{equation}
where the covariant derivative
\[
\hat D= \gamma ^\mu D_\mu  \quad D_\mu =\partial _\mu -ig\frac{\lambda
^a}{2}G^a_\mu ,
\]
$G_{\mu \nu }^a$ is the gluon field strength tensor,
$m=diag(m_u,\,m_d,\,m_s)$ and the matrices $\lambda ^a$ are the
generators of the $SU(3)_c$ color group.
By construction
 the ${\cal L}_{QCD}$
is an invariant under local gauge $SU(3)_c$ transformations.
Contrary to QED this lagrangian describes self--interaction of the
massless color gluons and contains the factors trilinear and
quadrilinear over the gauge fields  $G_\mu ^\alpha $.

Chiral invariance and \index{Chiral invariance}
vector nature of QCD
impose the important constraints on spin observables.
 Current quarks \index{Current quarks}
entering the QCD lagrangian have  small masses and may
be considered as massless objects. This is a good approximation for
$u$-- and $d$--quarks and sometimes it is also used  for
$s$--quark. In this case of $N_f=3$
the QCD lagrangian is invariant   under the chiral
transformations of the chiral \index{Chiral group}
$SU(3)_L\times SU(3)_R$ group, i.e. it is invariant under
the global transformations
\[
\psi _L\rightarrow L\psi _L, \quad \psi _R\rightarrow R\psi _R,
\]
where $L$ and $R$ stand for the $SU(3)$ transformations and
\[
\psi _R=\Gamma _R\psi ,\quad \psi _L=\Gamma _L\psi .
\]
The operators
\[
\Gamma _R=\frac{1}{2}(1+\gamma _5)\quad \mbox{and}\quad \Gamma
_L=\frac{1}{2}(1-\gamma _5)
\]
 are the projection operators.
 $\psi _L$ and $\psi _R$ are referred as left and right
chirality componenets, i.e. the chirality \index{Chirality}
is the eigenvalue of the Dirac
$\gamma _5$ matrix:
\[
\gamma _5\psi _L=-\psi _L, \quad \gamma _5\psi _R=\psi _R.
\]
QCD interactions  are the same for
the left and right quarks in the chiral limit $m_q\rightarrow 0$:
\begin{equation}
\bar{\psi }\hat D \psi =\bar{\psi }_L\hat D \psi _L+\bar{\psi }_R \hat D \psi
_R\,.
\label{4.25}
\end{equation}

As a result the left (right)--handed massless
particles will always
stay left (right) handed ones. Since
all the quarks are massless and have  the same QCD coupling, there
exists a separate $SU(3)$ flavor  invariance in right and left
worlds. This means invariance
 under transformation from $SU(3)_L\times SU(3)_R$ group.

So,  perturbative QCD deals with interactions at short
 distances and perturbative vacuum (invariant under the chiral
transformations). \index{Bare vacuum}

For massless quarks chirality and helicity coincide:
\begin{equation}
\psi _{1/2}=\psi _R,\quad \psi _{-1/2}=\psi _L.
\end{equation}
If the quarks have a non--zero masses the above operators $\Gamma _L$ and
$\Gamma _R$ do not yield the helicity precisely. Small mass term in the
QCD lagrangian leads also to explicit chiral symmetry breaking.
In that case chirality and helicity are equal approximately
 at high energies, namely:
\begin{equation}
\psi _{1/2}=\psi _R+O\left(\frac{m}{\sqrt{\hat{s}}}\right)\psi _L,\quad
\psi _{-1/2}=\psi _L+O\left(\frac{m}{\sqrt{\hat{s}}}\right)\psi
_R\,,\label{4.27}
\end{equation}
where indices
 $\pm 1/2$ denote the quark helicities.

Thus, any quark line entering Feynman diagram
\index{Feynman diagram} corresponding to the
lagrangian Eq. (\ref{4.22}) will emerge
with unaltered helicity since helicity flip amplitude is
proportional to current quark mass. The
quark helicity conservation is the most characteristic feature of
perturbative theory with vector coupling. For instance,
 tensor or pseudoscalar  exchange would ensure flip of
quark helicity.

In order to get a nonvanishing polarization it
is necessary
that the helicity flip amplitude being a  non--zero one and
in addition the phases of the helicity flip $F_f$ and non--flip
$F_{nf}$ amplitudes are to be different, since
\[
P\propto \mbox{Im}(F_{nf}F_f^*).
\]
In perturbative QCD the quark helicity flip amplitude is of the
order of $ m_q/\sqrt{\hat{s}}$ \cite{kpr}. Since
 the amplitudes are real
in the Born approximation, it is necessary at least to consider
the diagrams of
the fourth order in coupling constant $g$
 to get a non--zero imaginary part.
Thus, \index{Quark helicity flip}
quark helicity flip amplitude will be proportional
\begin{equation}
F^q_f\propto \frac{\alpha _sm_q}{\sqrt{\hat{s}}}F_{nf}^q\,,\label{4.28}
\end{equation}
and polarization has to be  vanishingly small in hard
interactions (where effective coupling constant is small):
\begin{equation}
P_q\propto \frac{\alpha _sm_q}{\sqrt{\hat{s}}}\label{4.29}
\end{equation}
due to  large value of $\sqrt{\hat{s}}\sim
p_{\perp}$ and small values of  $\alpha _s$ ¨ $m_q$,
where $m_q$ stands for mass of  current quark.
Even for the top quark with mass $m_t=140$ GeV the predicted value
of transverse polarization is equal to few percents
\cite{kanl}.

It should be noted that lattice calculations, low--energy
phenomenology and absence of parity doublets in particle spectrum
strongly indicate that
 in the real world
the chiral group $SU(3)_L\times SU(3)_R$ is broken down to $SU(3)_V$
with appearance of $N_f^2-1=8$ Goldstone bosons ($\pi $, K and
$\eta $). \index{Goldstone bosons} The dynamical realization of the
chiral symmetry breaking in the Nambu--Jona-Lasinio model suggests
that the pion, for instance, should be
considered as a collective state with strong admixtures of multi
$q\bar q$--components.

Therefore, one should note here, that even when the quarks are massless,
chirality is not a symmetry of QCD, it is broken (hidden) by the vacuum
state, which can be imagined as a complex mixture of virtual quark
pairs and is not invariant under the chiral group transformations.
 This is essentially a non--perturbative effect.

Thus one could expect significant spin effects due to non--perturbative
dynamics. However, our primary goal here is to consider the perturbative
QCD predictions.

\subsection{Helicity Properties of Exclusive Processes}
Convertion of conclusions at the constituent level into the
predictions the predictions for hadron
 is very complicated
problem in exclusive processes. The general
recipe is the use of the factorization theorems for exclusive and
inclusive processes \cite{fact}.

It should be noted that even in the leading order, factorization
is not a trivial procedure for hadronic processes.
Let us turn first to the Brodsky--Lepage approach \cite{brl} to
factorization in exclusive processes.
The important feature of the approach
is  that each parton distribution amplitude is related to a single
hadron. This assumption allows to separate
 bound state dynamics (long distance interactions) from
perturbative dynamics of hard parton scattering.
The integration over all transverse momenta of the constituents
$[d^2k_{\perp}]$ in definition
of the distribution amplitude for exclusive processes
 projects hadron wave function onto the state
with  $L_z=0$ (see, \cite{brl} and, e.g. \cite{ttqcd},
\cite{ralsp}). Hence, hadron helicity is
equal to the sum of the valence quark helicities:
\begin{equation}
\sum_{i=1}^n\lambda _i=\lambda _h.\label{4.48}
\end{equation}
This equation is valid in the framework of the above approach
 in all orders in $\alpha _s(Q^2)$ and in the leading order in $1/Q$.

The amplitude  of constituent interactions
preserves the total helicity of the valence quarks with accuracy
up to vanishingly small terms of order of  $O(m_q/Q)$, where $m_q$ is
 current quark mass. Along with Eq.
(\ref{4.48}) the latter conclusion leads to the helicity
conservation law --- the
sums of hadron  helicities in the initial and
 final states are equal, i.e. for reaction
$A+B\rightarrow C+D$  \cite{brl}:
\begin{equation}
\lambda _A+\lambda _B=\lambda _C+\lambda _D.\label{4.49}
\end{equation}
Eq. (\ref{4.49}) provides important experimental consequences.
The helicity conservation rule results in vanishing of
the one--spin asymmetries for  hard exclusive processes.
For example, the analyzing power in elastic $pp$--scattering is
proportional to the amplitude
 $F_5$,
which describes the transition
\[
|1/2,1/2>\rightarrow |1/2,-1/2>
\]
between the initial state with helicity 1 and the final
state with helicity 0.
 Eq. (\ref{4.49}) is not satisfied then and
\[
F_5=0.
\]
The Brodsky--Lepage helicity conservation rule provides  unambiguous
predictions for the elastic $pp$--scattering  at
large angles. In the leading order:
\begin{equation}
A=P=A_{sl}=0,\quad \mbox{and}\quad  A_{nn}=-A_{ss} \label{0}.
\end{equation}

This prediction is in disagreement with the available
experimental data indicating increase of the analyzing
power with the momentum transfer.

However, the above hard scattering picture meets with certain complications
of the above hard scattering
picture. The most significant ones are associated with independent
scatterings of quarks \cite{land}.
Such a picture
 demands that all three separate scatterings rather than one are
to be hard. In the framework of this Landshoff mechanism
 quarks before the
collision are in the state with  relative transverse distance
 $\Delta b\sim 1$ Fermi and suffer independent scatterings at the
same angle. The relevant amplitude corresponds to the
independent quark scatterings.
For $pp$--scattering the independent scattering mechanism is
determined by the three independent processes of
 $qq$--scattering at the same angle
$\theta _i\sim \theta _{c.m}$. These three quark scatterings take place along
the normal to the scattering plane.
 Simple factorization of small and large
distances does not take place here.

Comprehensive studies of
the both hard scattering mechanisms for hard exclusive processes
indicate that the
quark independent scattering dominates at very high energies
\cite{mul}, \cite{bots}.

The quark independent scattering has no
rotational invariance and therefore involves  the states with
non--zero orbital momentum \cite{bots}, \cite{ralsp}.
Orbital momentum may be transformed into spin of the hadron at
long--distance evolution and therefore  helicity conservation will not take
place for the hadron scattering while it does for the quark scattering.
This approach  allow the hadron helicity
flip without flip of a quark helicity and
 provides a hope that the observed large spin
effects would be explained within perturbative QCD.

\subsection{Experiments to Study Spin Effects in Exclusive
Reactions}
In the nearest perspective the spin phenomena studies
are expected to be carried out at  the fixed target facilities.
 Of the primary interest here are the measurements of the spin
observables in the high--energy elastic scattering at high
$p_{\perp}^2$. Such measurements are planned, e.g. at
the UNK with the use of  internal  polarized  jet  target
(NEPTUN--A experiment) \index{NEPTUN--A experiment}
\cite{nepa} at the energy range 400 -- 3000  GeV.

General arguments and comparison
with experimental data show that power law fall--off of
differential cross--sections is  valid at
quite low momentum transfers. For example, fixed $\theta _{c.m.}$
scaling for elastic $pp$-scattering is in agreement with
experimental data starting at $\sqrt{s}=5$ GeV and
$\theta _{c.m.}\simeq 40^o$, i. e. for $p^2_{\perp}=2-3$
(GeV/c$)^2$.
The measurements of the  analyzing  power and the spin--spin
correlation
parameters at large $p^2_{\perp}$ values were performed in the 10 to 30
GeV laboratory energy region. The experiments unambiguously demonstrated that
the relations (\ref{0}) are violated \cite{krs}.
Large one--spin asymmetry $A$ observed at $p_L=28$ GeV/c and high
$p^2_{\perp}$ reveals pronounced tendency (Fig. 1) to rise with $p^2_{\perp}$
and it reaches 24 $\%$ at $p^2_{\perp}=6.5$ (GeV/c$)^2$.
\begin{figure}
\vspace{7in}
\caption{The analyzing power, $A$ is plotted against $p^2_{\perp}$
for polarized $pp$ elastic scattering at $24$ and $28$ GeV/c.}
\end{figure}

The main goal of the NEPTUN--A experiment is to determine if the
unexpected large values of $A$  found in the proton--proton
elastic scattering at the AGS persist to the energies of hundreds
GeV. Besides that the NEPTUN program \index{NEPTUN
program}stipulates the studies of \cite{nep}:
\begin{itemize}
\item
 asymmetries in the inclusive production of  charged  and  neutral
mesons, as well as the photon production in  hard  interactions  on  a
polarized target;
\item
 asymmetries in the production of jets and lepton pairs;
\item
 polarization in the inclusive production of  hyperons,  the  spin
transfer parameter.
\end{itemize}
This program should allow to make
definite conclusion on validity of perturbative QCD due to
exploration of the new energy region where contributions
of the higher twists are expected to be small at high
$p^2_{\perp}$.
The experiment will also collect the data on the nucleon spin
structure. We will discuss these issue more thoroughly below.
The experiment is intended to run
with the use of the unpolarized proton beam at the UNK starting
from energy 400 GeV and later at 3 TeV.

It seems important to clarify what kind of hard scattering
(independent quark scattering, Brodsky--Lepage mechanism or
non--perturbative quark interaction)
 gives the main contribution to hard scattering. It would be desirable
to answer when the experiments would reveal analyzing power
at high energies. Valuable information
in that direction could be gained from nuclear target experiments.
In these experiments a nuclear target serves like a filter to
eliminate components of hadron wave function with large
transverse separation between the quarks \cite{coltr}. The
measurements of nuclear analyzing power for such components should
reveal attenuation of the analyzing power to zero  for a nucleus
with critical atomic number. The analyzing power would stay zero
beyond that number. and above that number analyzing power
would stay zero\cite{coltr}. However, if the analyzing power still
persists for nuclei with atomic numbers above the critical one
 then one
should arrive
to conclusion on a non--perturbative origin of spin effects in
hard scattering. Thus, the nuclear target experiments are very
useful to discriminate various hard scattering models.

An interesting proposal has been done recently by the SPIN
collaboration. \index{SPIN collaboration}
 It was suggested to upgrade the Tevatron (FNAL) to get
 the accelerated polarized proton beam \cite{spin}.
The proposal also includes  acceleration of polarized protons to 120
and 150 GeV in the \index{Fermilab Main Injector}
Fermilab Main Injector and the study of the
spin--spin parameter $A_{nn}$ and the spin--orbit asymmetry $A$ in
proton--proton elastic scattering at high values of $p_{\perp}^2$.
The design of the polarized proton capability for the
\index{Tevatron--Collider} Tevatron--Collider has been commissioned
as a further development
to study wide range of spin phenomena at Fermilab.

The primary physics goal of the SPIN is to determine if
 large spin--spin forces revealed by the previous $A_{nn}$ measurements
persist at the energies as high as 120 GeV. There is also possibility to
measure one--spin asymmetry $A$. In that respect this study would
extend the NEPTUN--A program at the UNK to cover the energy range
close to 100 GeV.

Polarized proton beam of high intensity is an essential
tool to measure the analyzing  power and spin--spin correlation
parameters at high--$p^2_{\perp}$ values. It seems important to
proceed with these measurements in the region of hundreds GeV
first to trace the experimental discoveries of the tens GeV
region. It is also important to carry out measurements of spin
parameters in the wide range of hard exclusive reactions in the
region of the tens of GeV (experiment EVA at Brookhaven).

\subsection{Tests of QCD in Inclusive Processes}
There are wide  opportunities for the use of accelerated
polarized beam at the Tevatron--Collider.
Results from SLAC and CERN (EMC) on the proton spin structure of
the proton together with the significant spin effects observed in hard
elastic and inclusive processes (Figs. 1--3) show that
further experimental studies
 at TeV colliders should be given a high priority.
Such experiments at the Tevatron--Collider with a
 polarized proton beam will both probe the fundamental couplings
of the lagrangian and
investigate the proton spin structure.
\begin{figure}
\vspace{7in}
\caption{Spin--spin correlation parameter $A_{nn}$
  for fixed c. m. s.
angle ($90^0$)  plotted against $p_{lab}$}
\end{figure}
\begin{figure}
\vspace{7in}
\caption{The polarization of $\Lambda $--hyperons produced by $400$ GeV protons
 plotted against $p_{\perp}$.}
\end{figure}

The polarized proton--antiproton collisions at $\sqrt{s}=2$
TeV would provide the new unique opportunities to study spin phenomena
at the highest energy and allow to test the Standard Model as well as
to search for the new physics beyond the Standard Model. Moreover
in the fixed target mode with the use of polarized target the
additional availability of polarized proton beam will allow to measure a
broad spectrum of the two--spin asymmetries at $p_L=1$ TeV.

One of the main goals of these studies should be the tests of QCD.
If the spin densities are known, the factorization theorem
allows to calculate the cross sections of hard processes and
corresponding one-- or two--spin asymmetries
 in the leading order for the respective
cross section differencies (constituent level):
\begin{itemize}
\item
for one--spin asymmetry, when the particle $A$ is polarized
\begin{equation}
A_Y\sigma ^{A+B\rightarrow C+X}=\sum_{a,b,c,d}\int \Delta _YG_A^a G_B^bD_c^C
\Delta _Y\sigma ^{a+b\rightarrow c+d},
\end{equation}
\item
for two--spin asymmetry,  when  the  particles  $A$  and  $B$  are
polarized
\begin{equation}
A_{YY}\sigma ^{A+B\rightarrow C+X}=\sum_{a,b,c,d}\int \Delta _YG_A^a\Delta
_YG_B^bD_c^C
\Delta _Y\sigma ^{a+b\rightarrow c+d},
\end{equation}
\item
for two--spin asymmetry, when  the   particle  $A$  is polarized  and
spin state of the particle $C$ is measured
\begin{equation}
D_{YY}\sigma ^{A+B\rightarrow C+X}=\sum_{a,b,c,d}\int \Delta _YG_A^aG_B^b\Delta
_YD_c^C
\Delta _Y\sigma ^{a+b\rightarrow c+d}.
\end{equation}
\end{itemize}
Here $\Delta _Y$ denotes differences of the corresponding quantities for
the different orientations ($Y=N$ (transversal) or $L$
(longitudinal)) of
single or both spins for the initial or final particles. The function
$G^a_A$ is the density of constituent $a$ in the hadron $A$.

When there are no other vectors
measured
in experiment
 besides spin of one particle and
momenta of initial particles $A$ and $B$ and final particle $C$
then the parity
conservation of strong interactions implies, that one--spin
longitudinal asymmetries must be zero, since the spin is an axial
vector. Indeed, spin vector of the polarized particle should be
contracted with another axial vector. But the only axial vector in
the case  is a normal
to the scattering plane. Therefore, the only possible asymmetry is
connected with the transverse component of the initial particle spin
vector.

Since the factorization theorems are valid for the transversely
polarized hadrons the definite conclusion with respect to the
one--spin transverse asymmetry in the leading order may be obtained.
Because of vanishing asymmetry in a hard subprocess
\[
a\propto \alpha _s\frac{m_q}{\sqrt{s'}},\label{4.57}
\]
the one--spin transverse asymmetry  $A_{N}$ for the hadron process should
also vanish.
 For example, the asymmetry $A_{N}\simeq 0$ in the leading order
in the reactions
\[
\pi ^-+p_{\uparrow}\rightarrow \pi ^0+X
\quad
\mbox{and}
\quad
p_{\uparrow}+p\rightarrow \pi ^0+X
\]
 since at the constituent level $a\simeq 0$. This should be also
true for polarization of  $\Lambda $--hyperons in the process
\[
p+p\rightarrow \Lambda _{\uparrow}+X.
\]
$P_\Lambda \simeq 0$, since polarization of quark appears to be
$P_q\simeq 0$.
The experimental data do not follow these
predictions at least up to the transverse momentum values
 $p_{\perp}\leq 3.5$ GeV/$c$.

\subsubsection{Higher Twist Contributions}
One of the possible escapes is to assume that the
transverse momenta values close to
$p_{\perp}\simeq 4$ GeV/$c$ are too small the leading order
calculations to be reliable. In that case one tries to account for
the confinement related effects such as higher twists
contributions, presence of  diquarks in the hadrons, interactions
in the final state, etc. At present time all these effects may be
taken into account
via  model approaches only.
However, considerations of the higher twist contributions in the framework
of perturbative QCD require
a minimal number of model assumptions. For the first time such
contributions
were used for the transverse asymmetry considerations in
paper \cite{et}.
In the recent paper \cite{qs2} the  higher twist effect
 were considered with the use of the generalized factorization theorem
\cite{fact}.
There was calculated the one--spin transverse spin
asymmetry in the leading order
for the direct--photon production process
\[
p_{\uparrow}+p\rightarrow \gamma +X
\]
at large transverse momentum of the photon.
 These calculations are based on the
consideration of the twist--3 matrix element $T(x,s_{T})$ of quark
and gluon fields. \index{Twist--3 matrix element}
The account for the quantity $T(x,s_{T})$ \cite{qs2} implies
presence of the correlations between the quark fields and the
strength of gluonic fields and thus assumes
the interaction between quarks and color fields .
Therefore, simple parton interpretation disappears.
Here an important role belongs to the orbital angular momentum.
 The asymmetry
\[
A(s_{T},x_F,l_{\perp})= E_l\frac{d\Delta \sigma _{T}}{d^3l}/
E_l\frac{d\sigma }{d^3l},
\]
\[
\Delta \sigma _{T}\equiv \frac{1}{2}[\sigma (s_{T},l)-\sigma (-s_{T},l)],
\]
where $l$ is the photon momentum, may be expressed in terms of the
parton densities, the matrix element $T(x,s_{T})$ and the hard scattering
functions accordingly to the generalized factorization theorem
\cite{fact}. The model assumptions are needed for the
$T(x,s_{T})$ as well as for quark and gluon spin densities. It was
proposed \cite{qs2} to model the
quantity $T(x,s_{T})$
with a mass scale times dimensional functions of $x$:
\[
T(x,s_{T})\cong 0.2F_2(x)/x\quad \mbox{GeV}
\]
or
\[
T(x,s_{T})\cong 0.2F_2(x)\quad \mbox{GeV},
\]
where $0.2$ GeV is a mass scale and $F_2$ is the structure
function measured in the unpolarized deep--inelastic scattering.
For the both choices of $T(x,s_{T})$ the asymmetry rises to over
20\% as $x_F$ approaches 0.8 (for $\sqrt{s}=30$ GeV and
$l_{\perp}$=4 GeV/c). This effect originates from the variation
 with $x$ of the function $x\frac{\partial }{\partial x}T(x,s_{T})$ entering
the
expression for the cross--section.

It  was  shown   \cite{jic}   that   at   large   negative   $x_F$
values the twist--three pure  gluon correlations become dominant in
the direct photon  production  on  the transversely polarized
nucleon.

Of course, the essential issue to obtain
a non--vanishing one--spin asymmetries is the presence of non--zero
relative phase between the helicity flip and non--flip amplitudes. The
recently proposed new QCD production mechanism for hard processes
at large $x$ \cite{hoy} leads to the relative phase
due to imaginary part of the diagrams with intrinsic $q\bar
q$--pairs. The quarks in the initial Fock states with such pairs
are localized in the region of a small transverse size and
therefore they have large transverse momenta.

These  results are promising ones and allow  to hope that
the
large transverse
asymmetries at large $x_F$
observed at the existing facilities
(at moderate energies) can be explained within QCD as the
higher--twist effects.
However, some  model assumptions are still needed, in particular, for
the choice of the form of higher twist contributions. Thus, the experimental
 studies of spin effects at large $x_F$ values will help to clarify
the particular dependence of the higher twists
expected to contribute at moderate energies and $Q^2$ values.

\subsubsection{Experiments to Test QCD in Inclusive Reactions}
However, the general statement, that in the leading order
the one--spin transverse asymmetries at the constituent level are
vanishingly  small  due  to  vector  type  of the  gluon   interaction
 in QCD is unchanged.
 Therefore, once the energy and transverse momentum
are high enough to rely  on the perturbative  expansion  in  QCD  and
 to neglect the higher--twist contributions  one should expect
$A_N=0$.
Any violation of this prediction would indicate the necessity of serious
modifications  of perturbative   QCD   and   account   for
non--perturbative effects.
Such helicity conservation violation could be attributed to
the chiral symmetry breakdown by the physical vacuum, since the spin
properties are intimately tied to the chiral structure of the
theory.

The energy independence of large polarization of $\Lambda $-hyperons
observed  in the region between 12 and 2000 GeV strongly indicates
that spin  effects persist at high energies. The use of the polarized
target and polarized beam will allow to measure the two--spin
correlations in the hyperon production at TeV energies and to
reveal the underlying mechanism leading to this still unexplained
result.

Moreover, the measurements of the longitudinal asymmetries will
provide the data on the production mechanism and on
the hadron spin structure. It seems very promising  to consider
the reactions with weakly decaying baryons in the final state such as
\begin{equation}
p_{\rightarrow}+\bar{p}\rightarrow \Lambda _{\rightarrow}+X
\end{equation}
and to measure the parameter
\begin{equation}
D_{LL}=\frac{\left. E_C\frac{d\sigma
}{d^3p_C}\right|_{^\rightarrow_\rightarrow}
-\left. E_C\frac{d\sigma }{d^3p_C}\right|_{^\rightarrow_\leftarrow}}
{\left. E_C\frac{d\sigma }{d^3p_C}\right|_{^\rightarrow_\rightarrow}
+\left. E_C\frac{d\sigma }{d^3p_C}\right|_{^\rightarrow_\leftarrow}}
\end{equation}
where the lower arrow denotes the longitudinal polarization
 of $\Lambda $--hyperon in the final state.
For such a reaction, the
subprocess $q\bar{q}\rightarrow s\bar{s}$
will dominate  at high $x_{\perp}$ values, while
the gluon annihilation $gg\rightarrow s\bar{s}$ will be important
 at small $x_{\perp}$ values. Therefore this process is quite
sensitive to the quark and gluon polarizations inside a proton.
Asymmetries at the constituent level are calculable in the
framework of perturbative QCD \cite{craig}; therefore
measurements of the  asymmetries $D_{LL}$ will provide the data
on quark and gluon spin densities. The theoretical estimations
provide  rather significant  values  for  $D_{LL}$  at the level
of  $50\%$ \cite{bsrt} that cause no problems in the $D_{LL}$
experimental measurements.
The measurements of $D_{LL}$ in the fragmentation region at large
$x_F$ seem to be interesting also from the
point of view of the polarization of strange sea and strangeness
content of a nucleon. Strange sea
may have a large negative polarization accordingly to the SLAC and
EMC  data interpretations and it should be revealed in $D_{LL}$
analysis.

The problem of the nucleon spin structure studies  will be considered
 further.

 The experimental results on elastic
scattering as well as results on the inclusive processes  support  the energy
persistence of spin effects.
As it was mentioned there is a chance to relate these spin effects to
 manifestations of the higher--twist terms (quark--gluon and
gluon--gluon correlations) \cite{qs2} \cite{jic}.
Therefore, the experimental   measurements   of $A_N$  in   the relevant
kinematical region  would reveal  the
size of the quark--gluon and pure gluonic correlations.
However, it could be expected that at the Tevatron--Collider
energy $\sqrt{s}=2$ TeV the higher--twist terms can  be safely
neglected at large $p_{\perp}$ values.  At such energies the $A_N$
measurements
 should  allow a clear test of perturbative QCD  in such processes
as:
\begin{equation}
p_{\uparrow}+\bar{p}\rightarrow \pi ,\gamma +X,
\end{equation}
or
\begin{equation}
p_{\uparrow}+\bar{p}\rightarrow jet+X
\end{equation}
at large $p_{\perp}$ values.

The measurements of $A_N$ in the above processes  would also test
the  chiral structure of the  effective
lagrangian approaches in the framework of non--perturbative
realization of  QCD.

\subsection{Nonperturbative Treatment of Spin Effects}

The data on the unexpected spin effects at small distances stimulated
development of the model approaches. There was proposed a number
of models to describe the spin phenomena. Some of the models use
the methods and ideas of QCD, but the other ones are only inspired by
QCD.
It should be emphasized that
description of the spin effects at large angles and in
particular of the behavior of the one--spin asymmetries
represent a complicated problem not only for
perturbative QCD but for the most of the QCD--based
models also. If the  present trends in  the experimental data
on the one--spin asymmetries will
persist at higher energies it could suggest a
non--perturbative origin of spin dynamics and will certainly initiate
the further development of the corresponding models.

 As it was already mentioned
there are two non--perturbative phenomena  implied by
QCD, namely, the confinement and spontaneous chiral symmetry
breakdown. As it currently believed, these phenomena are
characterized by the different scales.

It seems, that the crucial role under study of the spin effects belongs to the
chiral symmetry breaking.
The chiral symmetry breaking means that the chiral
symmetry of the QCD lagrangian is hidden by the complicated structure
of the QCD vacuum. Due to the chiral symmetry breakdown the hadron
structure already at the distance of tenth of Fermi diverges from
the parton model picture. \index{Chiral symmetry breaking}
The chiral symmetry breaking results in  generation of quark masses
comparable with the hadron mass scale. Therefore a hadron can
be  represented as a loosely bounded system of the constituent
quarks. The above observations related to the hadron structure lead to
understanding of several regularities observed in hadron interactions
at large distances. This picture provides also  reasonable  values  for
the static characteristics of  hadrons, for instance, for their magnetic
moments \cite{pondrom}.

Generally speaking the significant spin effects  observed
in hard processes over a long time pointed  out a non--perturbative  nature
of  hadron  dynamics, in particular, at rather short
distances.  The non--perturbative  dynamics at short
distances has been also revealed under the hadron structure
studies in deep--inelastic scattering. The  chiral  models appear to
provide
a transparent explanation  of the results of
the spin  structure
function $g_1(x)$ measurements   in deep--inelastic scattering of
polarized  muons on polarized proton target \cite{emc}. We return
to that important issue in the next section.

The usefulness of  the  models  based on the  chiral  symmetry  breaking
naturally leads  to the aspiration  to extend the
domain  of  their applicability to the region of small
distances \cite{shuryak}.

The attempts to incorporate some of the chiral model
ideas in a context of the unified approach to
a simultaneous description of hadron scattering both at large and
small distances were made recently. One should remember that the
theoretical approaches to description of hadron phenomena are
usually aimed for soft and hard interactions separately.
 The above mentioned
considerations are based
on the method of three--dimensional dynamical equations for
the scattering amplitude in Quantum Field Theory and on the current ideas
on the hadron structure and interaction of the constituents \cite{ttint}.
 This approach allows one to describe a number of spin effects.
 It also predicts:
\begin{itemize}
\item
non--vanishing one--spin asymmetry in elastic scattering at large
angles at asymptotical energies;
\item
oscillating dependence of the spin--spin correlation parameters in
elastic $pp$--scattering in the region of hundreds GeV;
\item
non--vanishing with energy and oscillating with $p_{\perp}$
polarization in the hyperon production at large $p_{\perp}$ values and $x_F$
close to 1;
\item
decrease with energy growth of the asymmetry in the pion production in the
central region.
\end{itemize}
Here we will not  discuss further here the model
approaches.
It is to be noted only that the models are very sensitive to the spin
observables and the new results will allow to discriminate various
approaches.
\section{Studies of Nucleon Spin Structure}
\subsection{Deep Inelastic Scattering and Spin Structure of
Nucleon}
Historically, the deep inelastic scattering of  electrons
indicated  first  the  presence  of  point--like  constituents   in
 a nucleon.
The spin dependent structure functions can be
measured in the deep inelastic scattering
 experiments with polarized lepton beam and a nucleon
target. These \index{Spin dependent structure functions}
functions allow one to extract the data on the nucleon's
parton distributions which characterize nucleon properties in hard
scattering processes. Measurements of the structure functions
have provided important results on the quark--gluon structure of
hadrons. The spin--independent quark and gluon distributions
were measured with high accuracy during twenty years. This is
not the case for the spin--dependent structure functions. The
longitudinal spin--dependent structure function has been
measured at SLAC and CERN. \index{SLAC} \index{CERN}

The two spin structure functions $G_1$ and $G_2$ enter the expression
for the
antisymmetric  part of the hadronic tensor
 $W_{\mu \nu }^{[A]}(p,q,s)$:
\begin{equation}
W_{\mu \nu }^{[A]}=\frac{1}{M}\varepsilon _{\mu \nu \lambda \sigma }q^\lambda
[M^2s^\sigma G_1(\nu ,Q^2)+
(p\cdot q s^\sigma -s\cdot q p^\sigma )G_2(\nu ,Q^2)].\label{7.12}
\end{equation}
In the Bjorken limit ( $\nu $ and $Q^2$ $\rightarrow \infty $, $x\rightarrow
$ const) the scaling is to be observed and
 the functions
 $G_1$ and $G_2$ should depend on $x$ only
up to the logarithmic corrections, i.e. they obey the relations:
\[
M^2\nu G_1(\nu ,Q^2)\rightarrow g_1(x),
\]
\begin{equation}
M\nu ^2G_2(\nu ,Q^2)\rightarrow g_2(x).\label{7.15}
\end{equation}

The recent data
obtained at CERN  by
the EMC \index{EMC} collaboration triggered a great number of
theoretical works
devoted to the problem of proton spin and resulted in the
extended interest in the future
experiments.
These data are in
disagreement with the expectations based on the
constituent quark model. Indeed, the
use of the $SU(6)$--symmetrical wave function of the proton
results in \begin{equation}
\Delta u=4/3,\quad \Delta d=-1/3,\quad \Delta s=0 \quad \mbox{and} \quad \sum_q
\Delta q=1.
\end{equation}

The  so--called  ``spin   crisis''   puzzle   emerged   from   the
experimentally obtained almost zero value for the sum
\[
\sum_q \Delta q\simeq 0
\]
 and
significant negative
polarization of strange quarks
\[
\Delta s=-0.18\pm0.07.
\]
The significant value of $\Delta s$  results also from the data on
elastic $\nu p$-- and $\bar \nu  p$--scattering \cite{ahr}. Thus,
extrapolation
to $q^2=0$
 of the measured values for the axial $Z^0$--$N$ formfactor  gives
\cite{kapl}:
\[
\Delta s=-0.15\pm0.08.
\]
 The above unexpected results stimulated a great theoretical
activity in the field. \index{Spin crisis}

In general, the SLAC and EMC experiments probe the hadron
structure at short distances, whereas the constituent quark model
is associated with the long range interactions. Of course, one
have to expect
that the structure of a nucleon at short distances differs from
its structure at long distances.  Nevertheless, the referred
experiments show up that the spin structure of nucleons is far from
trivial and deserves thorough consideration.

Shortly after the EMC analysis had been performed, the
perturbative QCD interpretation has been proposed. It is
based on account for anomalous gluon contribution to the
first moment of $g_1$ in the frameworks of parton model
\cite{pert}. In terms of the
operator product expansion this is related to the
 non--conser\-va\-tion of the axial--vector current
$J^5_{\mu 0}$ in QCD
 (even in the limit of massless quarks)
 due to the axial \index{Axial anomaly}
anomaly \cite{indu}:
\begin{equation}
\partial ^\mu J^5_{\mu 0}=2i\sum_q^{N_f} m_q\bar q \gamma _5 q +
N_f\frac{\alpha _s}{4\pi }G^a_{\mu \nu }\tilde G^a_{\mu \nu },
\end{equation}
where the dual $\tilde G$ is defined as
 $ \tilde G_{\mu \nu }=\frac{1}{2}\varepsilon _{\mu \nu \sigma \rho }G^{\sigma
\rho }$ and
$G_{\mu \nu }$  is the \index{Gluon field strength tensor}
gluon field strength tensor.
It was proposed to redefine the \index{Axial--vector current}
axial--vector current to ensure
its conservation in the limit of massless quarks:
\begin{equation}
\tilde{J}^5_{\mu 0}=J^5_{\mu 0}-K_\mu ,\label{7.22}
\end{equation}
where
\begin{equation}
K_\mu =\frac{\alpha _sN_f}{2\pi }\varepsilon _{\mu \nu \rho \sigma }G^{a\nu
}(\partial ^\rho G^{\rho \sigma }-
\frac{1}{3}gf^{abc}G^{b\rho }G^{c\sigma }).\label{7.23}
\end{equation}

This approach removes the anomaly contribution from the matrix
elements of the axial--vector current and provides a non--zero
contribution to the first moment of the structure function
 $g_1^p(x,Q^2)$. The matrix element of the difference $J^5-K$
is considered to provide the true  quark contribution $\sum_q \Delta q$
into the proton spin while the quantity  measured in experiment should
be replaced by $\tilde{\sum_q} \Delta  q $  according to the equation
\begin{equation}
\tilde{\sum_q }\Delta q=\sum_q \Delta q -N_f\frac{\alpha _s}{2\pi }\Delta
g.\label{7.43}
\end{equation}
If the used gauge does not imply appearance of the
unphysical degrees of freedom, then the
quantity $\Delta g$ could be interpreted  in terms of the gluon
densities with helicity parallel $g_+(x)$ and antiparallel $g_-(x)$  to the
nucleon helicity in the infinite momentum frame, i.e.:
\begin{equation}
\Delta g=\int_0^1dx[g_+(x)-g_-(x)].
\end{equation}
In such approach a photon can see the gluon helicity distribution
$\Delta g$ because of the axial anomaly resulting from non--conservation
of the singlet axial--vector current. The gluonic contribution
does not need to be a small correction because $\Delta g$ grows with
$Q^2$:
\[
\Delta g(Q^2)=\frac{\alpha _s(Q_0^2)}{\alpha _s(Q^2)}\Delta g(Q_0^2)
\]
 owing to the evolution equation \cite{kodr}.
It should be noted, however, that the operator $K_\mu $,
is not gauge invariant and hence it does not appear in the
 OPE.
The quantities entering Eq.\ref{7.43} depend
on the renormalization point.

In the sum rule for the proton spin
\begin{equation}
\frac{1}{2}=\frac{1}{2}\sum_q \Delta q+\Delta g+\langle L^q_z\rangle+
\langle L^g_z\rangle
\label{7.24}
\end{equation}
the individual components, except the first term on the right hand
side,  are not gauge invariants separately \cite{jffm}, \cite{kisp}.
Thus, the problems associated with the ambiguities in the
 gluon spin interpretation as matrix element of the current
$K_\mu $  make this interpretation controversial \cite{fort}.

 Moreover, to reconcile the EMC result with the constituent quark
model the large value for $\Delta g$ should be assumed:  $\Delta g\sim 5$.
Such value  can hardly be obtained in the models for the hadron
spin structure \cite{preparata}. In the case the large negative
orbital angular momentum ($\sim 5$) due to quarks and (or) gluons is
required to compensate the $\Delta g$. This orbital momentum value has to grow
logarithmically with $Q^2$ since $\Delta g\sim \alpha _s^{-1}$.
Thus, the anomalous gluon
interpretation
of the EMC result based on the hard gluonic contribution to the first
moment of $g_1$ leads to the new problems. Of course, nobody can
deny the possibility of final resolution of all these problems,
but under present circumstances, alternative approaches to the
problem of the proton spin also deserve consideration.

Thus, if $\Delta g=0$ or
we consider the values of $\Delta q$, extracted from the SLAC and EMC
data as the
true quark contributions to the proton spin, then
we should conclude, that the strangeness content of the proton, i.e. the
matrix element of axial--vector current $\langle
p|\bar{s}\gamma _\mu \gamma _5s|p\rangle$ has a large value.
In other words, the strange sea provide a significant negative
 contribution to the spin of proton.
The other indications on the strangeness content of the proton follow from
$\pi N$--scattering at low energies. In particular, from the value  of
the so--called \index{$\pi N$ sigma term}
$\pi N$ sigma term and the observed baryon masses  in
the first order over $SU(3)$ symmetry breaking
the value of
the  matrix  element   $\langle  p|\bar{s}s|p\rangle$   has   been
obtained \cite{mnhar}
\[
m_s\langle p|\bar{s}s|p\rangle = 334 \pm 132\,\, \mbox{MeV}.
\]
These results have important implications for the proton structure
and should stimulate further experiments to study
matrix elements of the strange quark operators.

\subsection{Experiments to Study Spin Structure of Nucleon}
The current experimental situation with the measurements
of
the spin structure functions is far from being complete.
Few independent experimental measurements of the
structure functions $g_1^{p,n}(x,Q^2)$
have been performed. There are
no data on  the second spin structure functions $g_2(x,Q^2)$ of
the proton and
neutron. In this situation the further experimental measurements
should play a crucial role.
\subsubsection{Deep Inelastic Scattering}
In deep inelastic scattering the  both structure functions
$g_1^{p,n}(x,Q^2)$ and $g_2^{p,n}(x,Q^2)$  may  be evolved
in experiments  with  the use of  longitudinally and transversely
polarized targets.  Such
measurements are planned at HERA in the framework of
HERMES project \cite{hermes}. \index{HERMES project}.
It is proposed to use the longitudinally polarized electron beam
and  polarized internal gas target. The
presumed types of the target materials are hydrogen, deuterium and
$^3$He. The electron beam energy should range between  30  and  35
GeV. The luminosity is expected to be varied from 3.5 to 30
$\times 10^{31}$ cm$^{-2}\cdot s^{-1}$. It will  allow  to obtain
the spin structure functions for protons and neutrons with the
high statistical precision.

This experiment will  also allow to discriminate the various
models  simulating  the  spin  structure  of  the  nucleon.  The
measurements will give the  contribution  of
valence  quarks  to  the  spin  of  the  nucleon,  since  under   the
simultaneous measurement of the proton and neutron spin  structure
functions, the contributions  of  sea  quarks  and  gluons  may  be
excluded. Being realized,  the  HERMES  project  will  also
provide the  possibility  to  check the
Bjorken  sum  rule \index{Bjorken  sum  rule}
derived in the framework of  QCD  with account for the isospin
invariance only and without  any additional assumptions.
This sum rule takes place
for the first moment of the difference $g_1^p-g_1^n$ \cite{bjor66}:
\begin{equation}
\int_0^1 [g_1^p(x)-g_1^n(x)]dx=\frac{1}{6}g_A(1-\frac{\alpha _s(Q)}{\pi })
\label{bjs}.
\end{equation}
Verification of Eq. \ref{bjs} will
provide an important test of Quantum Chromodynamics.

It is interesting that experiments SMC (Spin Muon Collaboration) at
CERN and E142 at SLAC have reported preliminary results on
the spin structure of the neutron. They allowed to test the Bjorken
sum rule for the first time. The SMC measurements with polarized
muon beam and polarized deuteron target support the original EMC
conclusions in particular on validity of the Bjorken sum rule.
However, E142  with polarized electron beam and helium--3 target
reveals a deviation from the Bjorken sum rule (3 sigma) and shows up
that more than a half of the proton's spin is due to quarks.

It is noteworthy that the experiments devoted  to  measurements of
the
spin structure functions are  now  recognized to be among the most
important ones and they are planned in many laboratories.

In the above discussions we focused on the deep inelastic
scattering on
 polarized protons and considered the structure function
 $g_1(x,Q^2)$
 related to the longitudinal polarization of a proton spin
with respect to its momentum, i.e.:
\begin{equation}
g_L(x,Q^2)=g_1(x,Q^2).
\end{equation}
In the parton model, $g_1$ measures the quark helicity density.
The second spin--dependent proton structure function
$g_2(x,Q^2)$  is related to transverse polarization of a nucleon
spin, i.e.:
\begin{equation}
g_{\perp}(x,Q^2)=g_1(x,Q^2)+g_2(x,Q^2).
\end{equation}
It has never been measured and
only few
theoretical results
 were obtained.
 The direct application of the parton model
orinally used for $g_{L}(x,Q^2)$ to
 the function $g_{\perp}(x,Q^2)$ results in the following
relation:
\begin{equation}
g_{\perp}(x,Q^2)=\frac{1}{2}\sum_q e_q^2\left(\frac{m_q}{xM}
\right)\Delta _{\perp}q(x,Q^2),\label{f}
\end{equation}
where $m_q$ is the quark mass, $M$ is the proton mass and
\[
\Delta _{\perp}q(x,Q^2)=q_{\uparrow}(x,Q^2)+\bar{q}_{\uparrow}(x,Q^2)
-q_{\downarrow}(x,Q^2)-\bar{q}_{\downarrow}(x,Q^2).
\]
 Eq. \ref{f} implies $g_{\perp}(x,Q^2)=0$ if the massless limit is
assumed $m_q\rightarrow 0$. This fact served as a basis for the
statement on the small value of the structure function $g_{\perp}(x,Q^2)$
and for the doubts in the applicability of the parton model to
interpretation of the second structure function and to
 description of the transversely polarized particles.

It should be noted first, that vanishing of
$g_{\perp}(x,Q^2)$ does not take place in the parton model
 as it was mentioned recently in Ref. \cite{ji}. Indeed, in this
model
because of the
relation $m_q=xM$ (where $M$ is the nucleon mass)
 the trivial zero--mass limit is inconsistent with the
non--zero mass of a nucleon in the rest frame.
Therefore in Eq. \ref{f} one should set $m_q=xM$. In this case
there is no reason for the smallness of the function
$g_{\perp}(x,Q^2)=0$. \index{Parton mass} Consideration of
 in the
simple parton model with non--interacting partons, i.e. when the
hadron is treated as a gas of free quarks, shows \cite{ansli} that
the spin density of quarks does not depend on the angle between
the nucleon polarization and its momentum. For that case
\begin{equation}
\Delta _{\perp}q(x)=\Delta q(x).\label{tr}
\end{equation}
Eq. \ref{tr}, however, is valid  for non--interacting
quarks only. In the more realistic models Eq. \ref{tr} should not take
place, in particular, $Q^2$--evolution would spoil it.

In general case one could rely on OPE.
The analysis based on the OPE does not depend on the type
(longitudinal or transversal) of the nucleon polarization.
Consideration of the second structure function
$g_{2}(x,Q^2)$
may be performed similar to the study of the function
$g_{1}(x,Q^2)$ \cite{ji}.  The  significant  difference,  however,
between, the two cases is that the function
 $g_1(x,Q^2)$ receives contribution from the twist--two operators
only, whereas $g_2(x,Q^2)$ gets contributions from the both
twist--two and twist--three operators simultaneously.
Note, that the simple partonic interpretation is
valid for the twist--two operator contribution. An account for
such contribution only provides the following relation for
the functions $g_1$ and $g_2$ \cite{ww}:
\begin{equation}
g_2(x,Q^2)=\int_x^1\frac{dy}{y}g_1(y,Q^2)-g_1(x,Q^2),\label{7.30}
\end{equation}
or in the other form
\begin{equation}
\int_x^1dx\,
x^{J-1}\left[\frac{J-1}{J}g_1(x,Q^2)+g_2(x,Q^2)\right]=0.
\label{7.31}
\end{equation}
The sum rule Eq. \ref{7.30} allows to calculate the function
 $g_2(x,Q^2)$ from the function $g_1(x',Q^2)$ at $x'\geq x$ in the
framework of the parton model with free on--shell partons.
However, there are no reasons to neglect the contributions of
the twist--three operators and therefore the function $g_2$
is to be represented as follows:
\begin{equation}
g_2(x,Q^2)=g_2(x,Q^2)^{[2]}+g_2(x,Q^2)^{[3]},\label{w}
\end{equation}
where the first term in the right hand side of Eq. \ref{w} is provided by
Eq. \ref{7.30}. The \index{Twist--three operator contributions}
twist--three operator contributions
$g_2(x,Q^2)^{[3]}$ depend on the effects of quark--gluon interactions and
quark masses. Due to the chiral symmetry spontaneous breakdown and
the confinement, these contributions should be large enough
 in any realistic model of a nucleon.

In particular,  the case of massive
off--shell quarks \index{Off--shell quarks}
 has been studied \cite{jrsr} in the framework of the
covariant parton model. \index{Covariant parton model}
 The following expressions for the
functions $g_1(x,Q^2)$ and $g_2(x,Q^2)$ were obtained:
\begin{equation}
g_1(x,Q^2)
=\frac{\pi x}{8}\int^{Q^2}dk^2dk_{\perp}^2
\left(1-\frac{k^2+k_{\perp}^2}{x^2M^2}\right)
\left(\frac{x^2M^2-k^2+k_{\perp}^2}{x^2M^2+k^2+k_{\perp}^2}\right)
\tilde{f}\left(x+\frac{k^2+k_{\perp}^2}{x^2M^2},k^2\right),
\end{equation}
\begin{equation}
g_{\perp}(x,Q^2)\equiv g_1(x,Q^2)+g_2(x,Q^2)=
\frac{\pi x}{8}\int^{Q^2}dk^2dk_{\perp}^2
\left(\frac{k_{\perp}^2}{x^2M^2}\right)
\tilde{f}\left(x+\frac{k^2+k_{\perp}^2}{x^2M^2},k^2\right),
\label{7.35}
\end{equation}
where the function  $\tilde{f}$  is determined by the quark
densities in the polarized nucleon.
{}From Eq. \ref{7.35} it follows that the function
$g_{\perp}(x,Q^2)$ is related to the mean transverse
parton momentum. \index{Transverse parton momentum}
The appearance of $k_{\perp}^2$ in the expressions for the
polarized structure functions shows that the deep inelastic
scattering processes with polarized
beam and target probe the features of the parton model different
from those in the deep inelastic scattering of unpolarized particles.

In the framework of operator product analysis
of the structure function, the
sum rule for the function $g_{2}(x,Q^2)$ can be obtained. This
sum rule was derived in Ref. \cite{buco} and is known as
Burkhardt--Cottingham sum rule \index{Burkhardt--Cottingham sum
rule}
\begin{equation}
\int_0^1 g_2(x)=0.\label{bc}
\end{equation}

It should be stressed that in the parton model the structure
function $g_1$ measures the quark helicity distribution in a
longitudinally polarized nucleon, whereas in transversely
polarized nucleon the structure function $g_{\perp}$ measures the
 average transverse spin for quarks and $g_{\perp}$ is related to the
quark--gluon interactions, i.e. to the higher twist terms.

The estimation of importance of the twist--three operators in the
framework of MIT bag model has been done in \cite{ji}. As it
was mentioned there are two types of the twist--three operators:
one is related to the quark gluon interactions and the other one
to the effect of quark masses. In the bag model the both types of
operators contribute to the function $g_2(x)$. These
contributions and the twist--two operator contribution are comparable
in their values.

Thus the experimental measurements of the structure function
$g_2(x)$ in
the deep inelastic scattering processes with polarized
particles
 are particularly important to reveal nonperturbative
effects
 related to the confinement and chiral
symmetry breaking. In the parton model the account for the
transverse spin--structure function assumes the study of
the off--shell partons and their transverse momenta.

\subsubsection{Direct Photon and Jet Production}
Besides the studies in deep inelastic scattering,
the spin structure of the  proton as well as the  role
of the gluon and sea components of spin, may also  be  investigated  in
the hadronic collisions. The hadronic processes allow, in
particular, to separate the quark and gluon contributions to the proton
spin. The EMC result interpretations are based on the hard
gluon  or sea quark
contributions  to the structure function $g_1$.
The first interpretation imply a large gluon contribution which in
principle should easily be detected. The sea quark interpretation
may be checked in the Drell--Yan processes.  The availability of
high--energy colliders with polarized proton beams would allow
experimenters to carry a number of studies sensitive to the
gluon and sea polarization. There are plans
 to have  polarized beams at the colliders \cite{ssc85}.
We discuss here some of the  experimental opportunities in this
connection.

In principle it is expected that the gluons and sea quarks are
equally important sources of the proton spin.

There are few  processes which allow to obtain
the data on the contribution of gluons to the proton spin.
It is production of the direct photons with large  \index{Direct
photons}
$p_{\perp}$ values in collision of the longitudinally polarized
protons:
\begin{equation}
p_\rightarrow+p_\rightarrow\rightarrow \gamma +X.\label{gm}
\end{equation}
Eq. \ref{gm} is one of the cleanest reactions to study perturbative
QCD, since the photon originates from the hard subprocess and
therefore there is no other photons resulting from the fragmentation
process.
The dominant contribution to the direct photon production  is due
to  the \index{Quark--gluon Compton subprocess}
quark--gluon Compton subprocess:
\begin{equation}
q+g\rightarrow \gamma +q.
\end{equation}
The  asymmetry is given by the expression:
\begin{equation}
a_{LL}=\frac{{s'}^2-{t'}^2}{{s'}^2+{t'}^2} \quad .
\end{equation}
The corresponding asymmetry at a hadron level, $A_{LL}$, is
determined by the two--spin asymmetry $a_{LL}$ and the longitudinal spin
densities of the
quarks $\Delta q(x)$ and gluons $\Delta g(x)$ in polarized proton:
\begin{equation}
A_{LL}=\frac{\sum_q\int [dx]\Delta q(x_1)\Delta g(x_2)
\frac{{s'}^2-{t'}^2}{{s'}^2+{t'}^2}\frac{d\sigma '}{dt'}+
\Delta g(x_1)\Delta q(x_2)\frac{{s'}^2-{u'}^2}{{s'}^2+{u'}^2}
\frac{d\sigma '}{du'}}
{\sum_q\int [dx]q(x_1)g(x_2)
\frac{d\sigma '}{dt'}+
g(x_1)q(x_2)
\frac{d\sigma '}{du'}}.\label{g}
\end{equation}
In Eq. \ref{g} the quantities $d\sigma '/dt'$ and
$d\sigma '/du'$ are the known unpolarized differential cross--sections
of the quark--gluon Compton subprocess.
Since contribution of this subprocess to
the asymmetry $A_{LL}$ is  proportional to the gluon polarization
$\Delta g$, the measurements of this quantity  would allow to extract
the data on
the gluon spin density provided
the  quark densities are known from the deep inelastic scattering
processes.

The direct photons can also  be produced in the quark annihilation
subprocess
\begin{equation}
q+\bar q\rightarrow \gamma +g.
\end{equation}
This subprocess is sensitive to the
sea polarization.\index{Sea polarization}
The respective asymmetry is $a_{LL}=-1$.
The statement that
the direct photon production is a good probe of the
gluon density is \index{Gluon density}
based on the observation that numerically the Compton contribution
to proton--proton interactions is  much larger than the
annihilation
contribution. It was recently shown that this is true for the
 case of polarized proton--proton interactions also and, in
particular, for the  case of longitudinally polarized protons
\cite{berq}.

The measurements of the direct photon production under the collision
of polarized protons may be carried out either with the use of
 fixed target accelerators
or colliders with polarized beams.

The gluon component of the proton spin should also manifest  itself  in
the measurements of $A_{LL}^{jet}$ and  $\Delta \sigma _L^{jet}$  in  the  jet
production process:
\begin{equation}
p_\rightarrow+p_\rightarrow\rightarrow jet+X.
\end{equation}
Contribution of  gluons into the quantity $\Delta \sigma _L^{jet}$ has
the following form:
\begin{equation}
\Delta \sigma _L^{jet}(s)=\frac{\pi \alpha _s^2}{2s}\sum_{i,j}\int_{x_1x_2>\xi
}
dx_1dx_2\frac{\Delta g_i(x_1)}{x_1}\frac{\Delta g_j(x_2)}{x_2}
\Delta H_{ij}(z^0),
\end{equation}
where $\xi =4(p_{\perp }^0)^2/s,\quad z^0=(1-\xi /x_1x_2)^{1/2},\quad
p_{\perp }^0$ is the cutoff determined by the  boundary  of  the
hard region. The function
$\Delta H_{ij}$ stands for the integral of the cross section difference
of the  two--particle
hard subprocesses. The initial gluons are to be in the
corresponding helicity states.
Due to large values of the gluon spin densities at small $x$,
$\Delta \sigma _L^{jet}(s)$ grows dramatically with increase of energy if the
net spin carried out by the gluons is sizable. For example,
$\Delta \sigma _L^{jet}(s)$  can be as large as $26$ $\mu b$ and $57$ $\mu b$
at
$\sqrt{s}=100$ and 200 GeV, respectively, whereas it is less than
$1$ $\mu b$ if the gluons are unpolarized \cite{qiu}.
Of course, the behavior of the two--spin asymmetry
$A_{LL}^{jet}$ is also essentially different for the cases of
polarized and unpolarized gluons.

\subsubsection{Drell--Yan Processes and Polarization of Sea}
The quark spin densities can be obtained under study of the
Drell--Yan production
of the lepton pairs \index{Drell--Yan process}:
\begin{equation}
p_\rightarrow+p_\rightarrow\rightarrow \mu ^++\mu ^-+X.\label{P}
\end{equation}
and
\begin{equation}
p_\rightarrow+\bar p_\rightarrow\rightarrow \mu ^++\mu ^-+X.\label{A}
\end{equation}
The process Eq. \ref{P} with the longitudinally
polarized initial protons is extremely suitable to test the sea
polarization, since $A_{LL}=0$ unless the sea quarks are polarized.
The sign of $A_{LL}$ is opposite to that of the sea quark
polarization. The expression for $A_{LL}$
has the simple form \cite{clos}:
\begin{equation}
A_{LL}=-\frac{\sum_q e_q^2[\Delta q(x_a)\Delta \bar q(x_b)+(x_a\leftrightarrow
x_b)}{\sum_q e_q^2[q(x_a)\bar q(x_b)+(x_a\leftrightarrow x_b)},
\end{equation}
where $x_a=\frac{1}{2}[x_F+\sqrt{x_F^2+4s'/s}]$ and
$x_b=\frac{1}{2}[-x_F+\sqrt{x_F^2+4s'/s}]$.
Hence, even measurement of the sign of
$A_{LL}$ will provide information on the sea quark
polarization in the polarized hadron.
 On the other  hand the process Eq. \ref{A}
 can be used to
extract the polarization of valence quarks. One could expect
large asymmetries  in  this  process.  However,  the  experimental
difficulties in realization of such collisions are evident.

If the two polarized nucleons are available in the  initial  state
then one could  measure \index{Three--spin  parameters}
the three--spin  parameters
such as $(l,l,l,0)$ in the process:
\begin{equation}
p_{\rightarrow}+p_{\rightarrow}=\Lambda _{\rightarrow}+X,
\end{equation}
where as always polarization of $\Lambda $--hyperon is studied through its
decay
process.
Measurements of the three--spin correlation parameters would be
important for study of the hyperon production dynamics and
the strangeness content of the proton.

\subsubsection{$\chi $--Production Processes}
Number of proposals have been
made to study the longitudinal and transverse spin structures of
nucleons at the Relativistic Heavy Ion Collider  under
construction at BNL \cite{bunce}. \index{RHIC} \index{BNL}
Besides the direct photons, jets and the Drell--Yan processes, the
study of $J/\psi $, $\chi _2$ and $\chi _0$ production at RHIC has been
considered. The $\chi $--production processes are interesting from
the point of view of the experimental measurements of the gluon spin
density \cite{cort}. The longitudinal two--spin asymmetry $A_{LL}$
for that case has the form \cite{cort}, \cite{donch}:
\begin{equation}
A_{LL}=a_{LL}\left(\frac{\Delta g(M/\sqrt{s}, M^2)}{g(M/\sqrt{s},
M^2)}\right)^2,
\end{equation}
where $M$ is the invariant mass of the final state and $a_{LL}$ is
the asymmetry for the  subprocesses
\begin{equation}
g+g\rightarrow\chi _2 \quad (a_{LL}=-1)
\end{equation}
or
\begin{equation}
g+g\rightarrow\chi _0 \quad (a_{LL}=+1).
\end{equation}
Thus the asymmetry
of inclusive $\chi _2$ production is a good probe of the gluon spin
density. The possible experimental difficulties related to the
different production mechanism contributions can be probably avoided
\cite{donch}.

There are also proposals to study the parity violation
effects at RHIC in the decay $W^+\rightarrow e^++\nu $ \cite{tann}.

\subsubsection{One--Spin Asymmetry Probe of Parton Spin
Densities}
In the above measurements the both initial state protons
 are to be polarized.
It can be realized in the polarized
proton beam scattering on polarized target or in the collision of
the two polarized  proton  beams.  At  present  such  experimental
facilities are not available. The \index{Tertiary polarized proton
beam}
tertiary polarized proton beam
used at Fermilab is not appropriate for that purpose because of its
low  intensity.  Acceleration  of   polarized   protons at
Fermilab or at the UNK  and the use of polarized target would provide
the necessary luminosity for  such hard scattering experiments in
the future.

Interesting proposal to use the existing
experimental facilities to study the proton spin structure has been
done recently \cite{carwil} (see also \cite{rspi}).
There the following process was considered:
\begin{equation}
p_\rightarrow +p\rightarrow \mu ^++\mu ^-+X,\label{cw}
\end{equation}
with one longitudinally polarized proton in the  initial state
when the momenta of the both outgoing muons
are measured. In this case the longitudinal asymmetry
$A$ for the process
Eq. \ref{cw} will be proportional to the quantity
\[
 \langle\vec s \cdot
{\vec{q}\;}^+\times {\vec{q}\;}^-\rangle,
\]
 where $\vec s$ is the proton
spin and ${\vec{q}\;}^\pm$
denote the momenta of the outgoing muons in the c. m. s. of the
colliding protons.

The asymmetry $A$ gets a non--zero value only when the outgoing muon pair
has a non--zero transverse momentum $Q_{\perp}$
($Q_{\perp}=q^+_{\perp}+q^-_{\perp}$).
Since the one--spin asymmetry $A$  arises
 from the one--loop contributions,  it is
proportional to the QCD running coupling constant $\alpha _s$.
However, it was shown \index{Running coupling constant}
the coefficients of $\alpha _s$ that originate from the parton
subprocesses are not necessarily small and that makes reasonable the
experimental measurements at moderate values of
$Q^2$.
Although it is
supposed that perturbation theory already works, the running
coupling constant is not too small to eliminate the effect.

The two parton subrocesses provide the contribution to the muon pair
production:
\begin{eqnarray*}
& & q+\bar q\rightarrow \mu ^+\mu ^-+g,\\
& & g+q \rightarrow \mu ^+\mu ^-+q,
\end{eqnarray*}
where one of the incoming partons is longitudinally polarized.
The asymmetry at the parton level may reach $30 \%$ \cite{carwil}

It is useful to have predictions at the hadron
level in order to reveal the dependence of asymmetry on the gluon spin
 of the total proton spin and to clarify the experimental
feasibility of these measurements. It appeared that at hadron level the
corresponding
asymmetry has  significant values at $p_L=70$ and $400$ GeV/c and
depends strongly on the gluon contribution $\Delta g$ (Figs. 4, 5).
The asymmetries at higher energies are decreasing (Fig. 6). The
corresponding asymmetries in $p\bar p$ collisions are rather small
($\sim 2 - 4\%$) due to large value of the unpolarized
cross--section of this process.
 Contrary to $pp$--collisions
the larger values of $\Delta g$ in $\bar{p}p$--collisions lead to the
smaller asymmetries.  The above
results have weak dependence on the specific parameterization of
parton densities.  The details of such calculations will
be published elsewhere \cite{nad}.
\begin{figure}
\vspace{7in}
\caption{One--spin asymmetry in the muon pair production process
$p_\rightarrow +p\rightarrow \mu ^++\mu ^-+X$
 versus the energy of the virtual photon in the c. m. s. at
$Q_{\perp}^2=0.4$ (GeV/c$)^2$ at $p_L=70$ GeV/c}
\end{figure}
\begin{figure}
\vspace{7in}
\caption{One--spin asymmetry in the muon pair production process
$p_\rightarrow +p\rightarrow \mu ^++\mu ^-+X$  versus
the energy of the virtual photon in the c. m. s. at
$Q_{\perp}^2=3$ (GeV/c$)^2$ at $p_L=400$ GeV/c}
\end{figure}
\begin{figure}
\vspace{7in}
\caption{One--spin asymmetry in the muon pair production process
$p_\rightarrow +p\rightarrow \mu ^++\mu ^-+X$  versus
the energy of the virtual photon in the c. m. s. at
$Q_{\perp}^2=50$ (GeV/c$)^2$ at $p_L=3000$ GeV/c}
\end{figure}

Thus, the measurements of the one--spin asymmetries in the muon pair
production $A=
 \langle\vec s \cdot
{\vec{q}\;}^+\times {\vec{q}\;}^-\rangle,$ will allow one to study
the gluon contribution to the proton
spin. There is no need to have very high energies for such
measurements and the NEPTUN experiment with polarized target at
the UNK seems to be the relevant facility.

The idea of determining the
polarization of an outgoing parton through the hadron distribution
in the corresponding jet \index{Jet handedness}
 (jet handedness) \cite{hand} is in line with above
consideration. The notion
of handedness imply the measurements of the momenta of three
particles from the jet and construction of pseudovector $n^\mu $ with
the use of these momenta. When
contracted with the polarization pseudovector of the outgoing
parton, it provides a parity conserving term in the decay
amplitude.

Thus, required experimental setup in the above cases may consist of
unpolarized proton beam and  polarized target or polarized proton
beam colliding with unpolarized proton (or antiproton) beam or
target.

\subsection{Transverse Spin Structure of Nucleon}
New possibilities are opening for measurements
of the transverse spin densities of quarks if the transversely
polarized proton beams available.
Here the exceptional role belongs to the
Drell--Yan processes with transversely polarized beams:
\begin{equation}
p_{\uparrow}+p_{\uparrow}\rightarrow \mu ^++\mu ^-+X, \label{lp}
\end{equation}
Recently,
 a wide discussion of the experiments on
transverse asymmetries was stimulated by the new theoretical
results \cite{rsop}.
In our consideration below we follow the above reference.

Let us remind first the definitions of parton
momentum and spin densities in the infinite momentum frame.
In that frame the momentum $P$ and spin $S$ of a proton
are
\begin{equation}
P=p+\frac{M^2}{2}n,\quad P^2=M^2;
\end{equation}
\begin{equation}
 S^\mu =S\cdot np^\mu +S\cdot p
n^\mu +S_T^\mu ,\quad S^2=M^2,\quad P\cdot S=0
\end{equation}
where $n$ and $p$ are null vectors which obey  the following
equations: $n^2=p^2=0, \quad n^+=p^-=0, \quad n\cdot p=1$. Their
mass
dimensions are -1 and 1, respectively. In the particular case of
the target moving in the $\hat z$ direction:
$p=1/\sqrt{2}(\tilde P,0,0,\tilde P),\quad n=
1/\sqrt{2}(1/\tilde  P,0,0,-1/\tilde  P)$,  where  the   parameter
$\tilde P$ characterizes the reference frame.

General
form of the quark spin density \index{Quark spin density matrix}
matrix element in the leading order for the massless case $m_q=0$
is as follows:
\begin{equation}
P^q_{\alpha \beta }(x,\lambda ,S_{T})=\frac{x}{2}\gamma ^\mu P_\mu
[q(x)-h^q_{L}(x)\lambda \gamma _5+h^q_{T}(x)\gamma _5\gamma ^\nu S_{\nu
T}]_{\alpha \beta }, \label{dm}
\end{equation}
where $P^\mu $ is the proton momentum, $\lambda $ is its helicity,
$S^\mu _T$ stands for the proton transverse spin as it was defined above
and $\alpha $  and $\beta $ are the Dirac indices. The quantities $q(x)$,
$h^q_L(x)\equiv \Delta q(x)$   and   $h^q_T(x)$   denote   the    unpolarized,
longitudinally and
transversely polarized quark densities, respectively.
All of them enter the density matrix with the factor of
 the proton momentum and in this sense the longitudinal and transverse
spin densities are comparable.

 For the case of deep inelastic scattering on
longitudinally polarized proton the quantity $h^q_{L}(x)$
contributes to the structure function $g_1$ in the  leading
order.
Contrary to $h^q_L(x)$ the transverse density $h^q_T(x)$ cannot be
measured in deep inelastic
scattering. The reason of such difference
  is in the different
behavior
of the two
parts of the quark spin density matrix
 under the chiral transformations. The transverse part of
 spin density matrix is odd: it commutes with the Dirac $\gamma _5$, while the
longitudinal contribution is
 anti--commuting with  $\gamma _5$ and thus is even.
The transverse density $h_T(x)$ measures the correlation  between  the
left--handed and the right--handed quarks.

Being chirally invariant the
electromagnetic
current cannot probe the transverse part of the density matrix.
Therefore it has been proposed to
consider the probe \cite{rsop}
 with different chiral properties than the electromagnetic current
has.
The relevant process is the Drell--Yan lepton
pair production Eq. \ref{lp} \index{Drell--Yan process}
where both incoming protons are polarized. The expression for the
cross--section of the process Eq. \ref{lp} in the one--photon approximation
is determined by the hadronic tensor $W_{\mu \nu }$, namely:
\begin{equation}
\frac{d\sigma }{dQ^2dyd\Omega }=\frac{\alpha ^2}{2(2\pi )^4sQ^2}\left(\delta
_{ij}-\frac{l_il_j}
{{\vec l}^2}\right)W_{ij},
\end{equation}
where $l$ is the lepton momentum, $Q^2$ is the mass squared of the
lepton pair and $y$ is its rapidity. The $d\Omega $ denotes the solid
angle of $\vec{l}$ in the dimuon rest frame where the right--hand
side is to be evaluated.
The hadron tensor $W_{\mu \nu }$ is defined by
\begin{equation}
W_{\mu \nu }\equiv e^{-2}\int d^4\xi \langle
P_AS_AP_BS_B|J_\mu (0)J_\nu (\xi )|P_AS_AP_BS_B\rangle, \label{ht}
\end{equation}
where $J_\mu $ is the electromagnetic current and $(P_AS_A)$ and
$(P_BS_B)$ are the momenta and spins of the hadrons $A$ and $B$.

In the one--photon  approximation the Drell--Yan process includes the
$q\bar q$--pair
annihilation described by the matrix element between the quark states:
\begin{equation}
\langle ks\bar k s'|J_\mu (0)J_\nu (\xi )|ks\bar ks'\rangle \rightarrow
\langle ks|q_\alpha (\zeta )J_\mu (0)J_\nu (\xi )\bar q_\beta (\chi )|ks
\rangle.
\end{equation}
Thus,  both the electromagnetic current and the antiquark field
are
the probes of the quark state.
A transversely polarized quark can be regarded as a superposition
of the longitudinally polarized states. Since
 antiquark interacts with quark conserving helicity
 and the antiquark field appears  in both (even and odd) chirality
combinations,
this field can be used to probe the even  (longitudinal)
 and odd  (transverse) parts of the quark spin density matrix.
 Of course, the above statement can be
reversed and the quark field can be considered as a probe of antiquark
state.

For the transverse spin density the expression in terms of the
matrix element of the bilocal operator has been obtained
\cite{shur}:
\begin{equation}
2S^\mu _Th^q_T(x,Q^2)=\frac{1}{2\pi }\int d\xi e^{i\xi x}\langle PS|\bar
q(0)i\sigma ^{\mu \nu }n_\nu \gamma _5q(\xi n)|PS\rangle,\label{pme}
\end{equation}
where $\sigma _{\mu \nu }\equiv \frac{i}{2}[\gamma _\mu ,\gamma _\nu ]$. Eq.
\ref{pme} was obtained
in the infinite momentum frame and it
   represents a generic form for the
field--theoretical definition of the parton densities. The expression
for $h^q_L(x,Q^2)$ has the similar form:
\begin{equation}
2\lambda h^q_L(x,Q^2)=\frac{1}{2\pi }\int d\xi e^{i\xi x}\langle PS|\bar
q(0)\gamma _5\gamma _\mu n^\mu q(\xi n)|PS\rangle.\label{pml}
\end{equation}

The OPE analysis of the function $h^q_T(x)$ determined as a
matrix element of the twist--two operator has been performed
\cite {shur}. As it was stressed,
 the quantity connected with the transverse quark density
$h^q_T(x)$, differs from the transverse spin  operator  $\Sigma _{\perp}$
by the $\gamma _0$ factor. Therefore,
 in general, for the case of off--shell
particle   the   eigenstate   of   the   transversity    operator
$\gamma _0\Sigma _{\perp}$ is not that of  the  transverse  spin
operator.  The transverse  spin  operator  does   not   commute   with
the free--quark Hamiltonian and depends on the underlying
dynamics \cite{kog}.
 Therefore it was proposed to name $h^q_T(x)$ as transversity.

 It  is  worth  to  note  here  that
$\Delta _{\perp}q(x)$ and $h^q_T(x)$ are different quantities
 in the parton model. The $\Delta _{\perp}q(x)$ measures an
average transverse spin of quarks  while the quantity  $h^q_T(x)$
measures the net number of quarks in the transversity eigenstate
\cite{rsop}, \cite{jip}.

 There were considered also the
higher--twist transverse structure functions which do not allow simple
partonic interpretations and are potentially useful for the studies of
nonperturbative effects in QCD.

It should be noted that quarks and antiquarks contribute
to $h^q_L(x)$ and $h^q_T(x)$ in different ways. The sum  rule  for
the function $h^q_T(x)$
\begin{equation}
\int_0^1dx[h^q_T(x)-h^{\bar q}_T(x)]=\delta q
\end{equation}
shows that contrary to $\Delta q$ the \index{Tensor charge}
``tensor charge'' $\delta q$ gets contributions from  the
valence quarks only.  There is no gluon
contribution to $h^q_T(x)$ due to the  angular  momentum  conservation
\cite{rsop}.
Contrary to $h^q_L$
it evolves with $Q^2$ uncoupled with the gluon density.
 Therefore the ambiguities related to the gluon
anomaly are not to be present here.

\subsubsection{Experiments to Measure the Transverse Spin
Quark Densities}
The quantities $h^q_T(x)$, $h^q_L(x)$ along with the unpolarized quark
density $q(x)$ determine the angular distribution of lepton pairs in
the Drell--Yan process \cite{cpr}:
\begin{eqnarray}
\frac{d\sigma }{dQ^2dyd\Omega }= & & \sum_q\frac{\alpha
^2e_q^2}{12Q^2s}\{[q^A(x_A){\bar
q}^B(x_B)-\lambda _A\lambda _Bh^{qA}_L(x_A){h}^{\bar q B}_L(x_B)]\times
 \nonumber \\
 & & (1+\cos^2\theta )+ S_{TA}S_{TB}h^{qA}_T(x_A) h^{\bar q
B}_T(x_B)\sin^2\theta \times
\nonumber \\
& & \cos (2\varphi -\varphi _A-\varphi _B)+ A\leftrightarrow B\},\label{ds}
\end{eqnarray}
where $e_q$ is the quark electric charge in units of the proton
charge. The $Q^2$--dependence is implied for the quark densities.
In Eq. \ref{ds}\, $\theta $\, is the polar angle and\, $\varphi $\, is the
azimuthal
angle both related to the beam direction;  $\varphi _A$ and $\varphi _B$ denote
the azimuthal angles of the transverse spin projections for the
nucleons $A$ and $B$. From the angular distribution Eq. \ref{ds}
an asymmetry $\Delta \sigma $ which depends on the transverse densities alone
can
 be obtained by means of integration over all the polar angles
 and subtraction of the integrals over the azimuthal angle $\varphi $:
\begin{equation}
\frac{d\Delta \sigma }{dQ^2dy}=\int_{\pi /4}^{3\pi /4}+\int_{5\pi /4}^{7\pi
/4}-
\int_{3\pi /4}^{5\pi /4}-\int_{-\pi /4}^{\pi /4}d\varphi \frac{d\sigma
}{dQ^2dyd\varphi },
\label{sb}
\end{equation}
where the spins of nucleons $S_{TA}$ and $S_{TB}$ are chosen to be oriented
along $\hat x$ axis ($\varphi _A=\varphi _B=0$).
The  quantity Eq. \ref{sb} in terms of the transverse quark
densities is
\begin{equation}
\frac{d\Delta \sigma }{dQ^2dy}=\frac{4}{9}\sum_q
\frac{\alpha ^2e^2_q}{Q^2s}h^{qA}_T(x_A)
h^{\bar q B}_T(x_B)S_{TA}S_{TB}+A\leftrightarrow B.
\end{equation}
Thus, the measurements of lepton pair angular distribution in
the Drell--Yan process \index{Drell--Yan process}
 will allow to determine the
 transverse spin  densities for quarks and antiquarks.
Until recently the transverse spin densities have been almost
ignored in deep inelastic scattering processes because of their
higher--twist
origin. Fortunately, the recent progress in that field made  clear
the necessity of the experimental studies of these densities in
the Drell--Yan processes. This is completely
unexplored domain of spin physics.

Besides the Drell--Yan processes there are several possibilities
to study the transverse quark densities  \cite{rsop}. One could
mention the process of $\pi ^+\pi ^+$--pair production
\begin{equation}
p_{\uparrow}+p_{\uparrow}\rightarrow \pi ^++\pi ^++X
\end{equation}
dominated  by
 the scattering of the two $u$--quarks
at large and opposite $p_{\perp}$ values
 and the semi--inclusive deep inelastic scattering reaction
\begin{equation}
e^-+p_{\uparrow}\rightarrow e^-+\Lambda _{\uparrow}+X.
\end{equation}
Since polarization of $\Lambda $ can be measured through its decay,
the measurements of the parameter $D_{NN}$ will allow to
restore the transverse spin densities of quarks.

In the recent paper \cite{hepp} it was shown that the transverse
spin of  quark initiating a jet can be measured through the
azimuthal dependence of pion pairs in the jet fragments.

There is also good chance to study the transverse spin densities of
quarks  in the direct $\gamma $ production in the polarized $p\bar p$
collisions.
 Contrary to the longitudinally polarized
case, only quark annihilation will contribute to the asymmetry in
the direct $\gamma $ production since gluons cannot
be polarized transversely in the spin--$1/2$ nucleon. At favorable conditions
the values of asymmetries can reach $10-20$ \% \cite{jip}.
The measurements of the two--spin asymmetries $A_{NN}$ in the
production of heavy quark pairs will also  supply the data on the
transverse spin quark densities.

\section{Spin Phenomena and Electroweak Interactions}
It should be noted that the spin  effects
expected under the searches and exploration  of  new
objects, should be observed in the presence of intensive  hadronic
processes. Therefore, to separate such  effects  one is to be
able to calculate the hadronic asymmetries. They
require in their turn
the knowledge of the corresponding quark and gluon spin densities.
Till now, the information  about  these  densities is
rather fragmentary.  The  perspectives  for  the
experimental studies  of  the  longitudinal  and  transverse  spin
densities will be discussed further. Their knowledge is also
vital
for variety of the  calculations in electroweak sector of the
Standard Model based on the $SU(2)_L\times U(1)_Y$ gauge
symmetry. There are four gauge bosons associated with this
symmetry: the well known $W^\pm$--, $Z^0$--bosons and photon. The
subscript $L$ reminds that the charge changing weak
interactions describe transitions between left--handed
fermions only (i.e. they violate parity conservation) while the Z
coupling to a fermion is a mixture of the
pure $(V-A)$ coupling of $SU(2)_L$ and the vector coupling of QED.

In electroweak interactions, where the parity  is  not  conserved,
there  could  be observed  significant  one--spin
longitudinal asymmetries as well as transverse ones.
Let us consider the case of $W^+$-- or
$W^-$--boson production.  The amplitudes of the corresponding
processes
\begin{eqnarray*}
& &p_\rightarrow + \bar p\rightarrow W^++X,\\
& &p_\rightarrow + \bar p\rightarrow W^-+X
\end{eqnarray*}
 may be calculated by means of account for
the contributions of the dominant fusion reactions only:
$u+\bar{d}\rightarrow W^+$ or
$\bar{u}+d \rightarrow W^-$. For the case of
longitudinally polarized proton the  asymmetry
at the level of the subprocess
  will  be maximal ($a_L=100\%$) since W is  the left--handed
current.
The asymmetry at the hadronic level
$A_L$
 will also become essential
 after $a_L$ is integrated with the quark spin densities
$\Delta q_i(x,Q^2)$. In the  above approximation to the Drell--Yan mechanism
 the asymmetry $A_L$ is an algebraic combination of the quark densities.
 The  values of $A_L$ in the above--mentioned
reactions can reach $60- 80\%$.   These asymmetries decrease very
slowly with the energy.
 Apart  from the independent test of electroweak sector of
the Standard Model, \index{Standard Model}
 the experimental studies of such a  dependence
allow  to get  information on the quark spin densities.

Due  to the chiral properties of $W$--boson the  asymmetry at the
level of the subprocess
$a_L$ will be universal for different pairs of quarks
$q_i$ and $q_j$ in the processes
\begin{equation}
q_i+\bar{q}_j\rightarrow W^{\pm}(M)\rightarrow
 X,\label{8.2}
\end{equation}
where the final state $X$
 contains   $W^{\pm}$  and  a  neutral
particle ($M$ is the invariant mass of the final state particles).

Similar  universality  will  characterize the asymmetries
$A_L$ in those hadronic processes where the contributions from
the subprocesses Eq. \ref{8.2} are dominant.
The following reactions are particular examples of such hadronic
reactions:
\begin{equation}
p+p\rightarrow W^{\pm}+\gamma ,\,W^{\pm}+Z^0,\,W^{\pm}+H^0.
\end{equation}
The details can be found in extensive review \cite{bsrt}.

The  minimal  extension  of  the  Standard Model consists in
introduction   of the
models with the left--right symmetry when the  parity violation  is
considered  as  a  low--energy  effect.  These   models imply
the    existence    of the right--handed    $W$--bosons.    The
asymmetries in   the production of
$W_R^+$-- and $W_R^-$--bosons  are directly  obtained  from  the
corresponding asymmetries in the production of left $W$--bosons by
means of reversal
of the asymmetry signs owing to the change of sign  of the axial  coupling
constant.

The collisions of polarized proton beam with the unpolarized one
could provide an additional information on a possible strong electroweak
sector. Recently spin asymmetries were calculated for that purpose
using BESS model (Breaking the Electroweak Symmetry Strongly)
\cite{casalb}. Here the spin asymmetries different from those predicted
by the Standard Model could indicate the presence of the strong
electroweak sector. BESS model predicts the vector resonances
(V--particles) which are bound states of a strongly interacting
sector. The measurements of the one--spin asymmetry $A_L$
in the processes
\[
p_\rightarrow+p(\bar p)\rightarrow l^+l^-+X
\]
could allow to distinguish $V^\pm$ bosons from $W_R$ boson in the
region where $q\bar q$--processes dominate. Thus, it allows one to
discriminate between the different models of electroweak
interactions.
\section{Search for New Particles and Spin Effects}
\subsection{Compositeness}
Polarization measurements may also be useful  for study of the
compositeness. Currently, a large number of  models  treat  quarks
and leptons as the composite particles.  The  interaction  between
the  new
constituents (preons) \index{preon} may generate the interactions
with  an arbitrary chiral structure which can violate the parity
conservation.

The both $A_L$ and $A_N$ measurements
 are  useful for search of the compositeness.
The simplest signal for the quark compositeness is
deviation of the jet production cross section at large transverse momenta
from the values predicted by perturbative QCD.
This deviation would arise from the new interaction  between
quarks induced by their composite structures
\begin{equation}
{\cal{L}}= {\cal{L}}_{QCD}+\eta _0\frac{g^2}{\Lambda _c^2}\bar{q}Aq\bar{q}Aq,
\end{equation}
where $\Lambda _c$ is the scale of the compositeness of the order of
the binding energy for preons (it is usually taken to be $\sim
1-2$ TeV), $A$ determines the Dirac structure
of interaction and depends on the details of particular composite
model. As it is generally believed the new interactions induce
a parity violating term in the lagrangian. \index{Composite model}
The above mentioned deviation from QCD--interactions  may be found
while measuring the one--spin longitudinal  asymmetry
 $A_L$ in hard hadronic processes. In virtue  of  the  parity
violation in the preon interaction, this  quantity  will  differ
from zero and become rather large at
$p_{\perp}\simeq  3-  4$  GeV/$c$.
Therefore, the large longitudinal asymmetries
in hadron interactions may occur due to manifestation of the preon
interactions.
\begin{figure}
\vspace{7in}
\caption{The asymmetry $A_L$ in reaction $pp\rightarrow jet+X$}
\end{figure}
Fig. 7   presents the calculations
\cite{txl} for the asymmetry $A_L$ in the  jet  production  in
hard  $pp$--interactions.

Noticeable effects will  be also observed
under production of direct photons or  lepton  pairs.
The compositeness  should enlarge   polarization   effects   in
these
reactions too.

 The important role of the parallel studies of the polarization and
the compositeness consists in providing the opportunities for the
choice of particular type of the interaction, i.e. the form of
$A$. Indeed, the different forms of $A$ such as $A=\gamma ^\mu (1-\gamma _5)/2$
or
$A=\gamma ^\mu $ provide
 almost the same predictions for the spin--averaged  cross
sections \cite{albr}. On the other hand
 predictions for the $A_L$ parameter
are essentially different for the above two options of $A$.

\subsection{Supersymmetry}
The SUSY theories predict the existence of a series of
new  particles.
The supersymmetric partners of ordinary  particles  are  known  to
differ by $1/2$ unit in their spins. Therefore, the
asymmetries  appearing
in  the  production  of  SUSY  particles \index{SUSY  particles}
  in the polarized   hadron
collisions, will differ from the corresponding asymmetries arising
in the production of ordinary particles.

Since light SUSY particles interact with ordinary  matter  weakly,
the production of supersymmetric  particles  is  characterized  by
the events with  the  missing energy--momentum.  It  is
difficult to interpret  such  events;  that  is  why  polarization
measurements may turn out to be quite  useful.  For  instance,  in  the
strong sector one should expect production of a large amount of
scalar quarks
(squarks, spin 0) and gluinos (spin 1/2). A characteristic feature
of this type events is  the  jet  production  and  the
``missing energy''.  Since in  strong  interactions the  parity  is
conserved one  should  consider  the
two--spin asymmetries
$A_{LL}$ as a relevant spin parameter. For the processes,
 where a pair of SUSY  particles
is produced  in  the  final  state,
the subprocess asymmetry
$a_{LL}^{ij}=-100\%$   for  the  case  of  massless  squarks   and
gluinos because of the helicity conservation.
\begin{figure}
\vspace{7in}
\caption{The asymmetry  $A_{LL}$  for  production   of the  ordinary
particle (\it a \rm) and supersymmetric particle (\it b\rm)}
\end{figure}
 As a result, the double asymmetry
$A_{LL}$ should be negative and have larger  values  than  in  the
case of the ordinary particles. To illustrate this statement Figs.
8,\it a \rm and
8,\it b \rm represent the predictions for the parameter
$A_{LL}$ in the two cases of the ordinary and supersymmetric particle
production respectively \cite{bsrt}.
Therefore,  the  most  typical  feature  of  the   SUSY   particle
production is appearance of the events with  specific  behavior  of
the polarization asymmetries.
\section{Spin Effects in $e^+e^-$--Collisions}
Electron--positron collisions are highly  interesting since
 the spin effects are  related
to the interactions at small distances. This is due to the fact that
$e^+e^-$--interactions are mediated by the electroweak
currents $(\gamma ,\,Z,\,W^{\pm},\ldots )$ which have  point--like
couplings and a simple spin structure.
$e^+e^-$--interactions allow also
high accuracy measurements.
The main goal
of any $e^+e^-$--collider program is the physics of
$Z^0$-- and $W^{\pm}$--bosons.
The interest in the respective polarization
 studies of these  processes  lies
mainly in the high energy region, where weak  interactions  become
especially important.

It should be noted that the   radiation  effects
lead to the appearance of a  natural  transverse  polarization  in
$e^+e^-$ storage rings
\begin{equation}
P_{\perp}(e^+)=-P_{\perp}(e^-)
\end{equation}
 which may exceed
$90\%$. It should be noted that the transverse polarization leads to
 a certain azimuthal dependence that was used to  determine
the  beam polarization and to verify the fermionic
nature of quarks.

In $e^+e^-$--collisions, the  colliding  particles  produce
a virtual intermediate state that, in accordance with the Standard
Model, is
either a photon or $Z^0$--boson.  In  this  case,  the
amplitudes   are   completely
defined by the coupling constants of the  gauge  bosons $(\gamma $ or
$Z^0$)  with
the final states. Therefore, spin  effects  in  these  reactions
depend on  the  initial  spins only,  whereas angular
asymmetries  are  related to direction of the beam
polarization.

The dependence of the $e^+e^-$--annihilation cross sections on  the
beam polarization appears due to interference of the  contributions
from the virtual $\gamma $ and
$Z^0$. Then, since  at  the  energies  higher  than  200  GeV  the
value of the contribution from weak  interactions  is  comparable
with  that  of
electromagnetic interactions, one   should
 expect significant polarization effects.
In this energy range the asymmetry $A_L$
\begin{equation}
A_L=\frac{\sigma (+-)-\sigma (-+)}{\sigma (+-)+\sigma (-+)}=
\frac{\sigma (+0)-\sigma (-0)}{\sigma (+0)+\sigma (-0)}=\frac{\sigma
(0+)-\sigma (0-)}{\sigma (0+)+\sigma (0-)}
\end{equation}
(by the signs $+,-$ and 0 we mean the longitudinal polarization
$P_L=+1,\,-1$ and 0, respectively) has the
following values \cite{bsrt} for the different processes:
\begin{eqnarray*}
e^+e^-  \rightarrow & e^+e^- & -\quad 7\%, \\
  \rightarrow & \nu \bar{\nu } & -\quad 15\%, \\
 \rightarrow & u\bar{u},c\bar{c},t\bar{t} & -\quad
34\%, \\
  \rightarrow &d\bar{d},s\bar{s},b\bar{b} & -\quad
62\%, \\
  \rightarrow &Z^0Z^0 & -\quad 32\%, \\
  \rightarrow &W^+W^- & -\quad 94\%.
\end{eqnarray*}
Verification of these values would provide an  important test  of
the Standard Model when the initial $Z$--fermion--antifermion
vertex
\begin{equation}
-ie\bar{u}_f\gamma ^\mu (a_f-b_f\gamma ^5)u_f
\end{equation}
is determined by  the  vector  and  axial  coupling  constants.
The corresponding values for leptons and quarks are given in the
 Table 1.
\begin{table}
\begin{center}
\begin{tabular}{||c|c|c|c|c||}\hline\hline
 & $\nu $ & $e$ & $u$ & $d$ \\ \hline\hline
 $2\sin\theta _Wa_f$ & $1$ & $-1+4\sin^2\theta _W$ & $1-\frac{8}{3}\sin^2\theta
_W$
&
$-1+\frac{4}{3}\sin^2\theta _W$\\ \hline
 $2\sin\theta _Wb_f$ & $1$ & $-1$ & $1$ & $-1$ \\ \hline\hline
\end{tabular}
\end{center}
\caption{Axial and  vector  coupling  constants  for  leptons  and
quarks, $\theta _W$ is the Weinberg angle}
\end{table}

Beyond the Standard Model, investigations  of  spin
effects  might  be  used  for  the  search  of  the  new
physics. Thus, for instance, the
$Z'$--boson, predicted in the \index{Grand  Unification Theories}
Grand  Unification Theories  should
manifest itself as  a  peak  in  the  $e^+e^-$--interaction  cross
section. If so, even out of the resonance region
 one would observe a  considerable  deviation
 of the asymmetry $A_L$  from the behavior predicted by the Standard
Model. The  measurements  of such
deviation would allow to extract the coupling constants for the
interactions of $Z'$ with different fermions.

In  $e^+e^-$--interactions the polarization effects  may
 be essential for  the search for the  SUSY
particles. Both longitudinally and
transversely polarized electron beams are equally important for
these purposes.

In conclusion it should be noted that spin physics experiments will
provide the crucial tests for the present theories and
undoubtedly will bring new unexpected results. This
justifies the efforts needed to perform the spin measurements and
to  create the new spin physics facilities.

We would like to thank A. D. Krisch, D. I. Patalakha, J. P. Ralston
and V. L. Solovianov for useful discussions, comments and
suggestions.

\end{document}